%% file: main.tex
\documentclass[sigconf]{acmart}
\usepackage{graphicx}
\usepackage{multirow}
\usepackage{hyperref}
\hypersetup{
    colorlinks=true,
    linkcolor=blue,
    filecolor=magenta,      
    urlcolor=cyan,
}

\AtBeginDocument{%
  }

\copyrightyear{2025}
\acmYear{2025}
\setcopyright{cc}
\setcctype{by}
\acmConference[WWW Companion '25]{Companion Proceedings of the ACM Web Conference 2025}{April 28-May 2, 2025}{Sydney, NSW, Australia}
\acmBooktitle{Companion Proceedings of the ACM Web Conference 2025 (WWW Companion '25), April 28-May 2, 2025, Sydney, NSW, Australia}
\acmDOI{10.1145/3701716.3715290}
\acmISBN{979-8-4007-1331-6/25/04}

\begin{document}

\title{RCAEval: A Benchmark for Root Cause Analysis of Microservice Systems with Telemetry Data}

\author{Luan Pham}
\email{luan.pham@rmit.edu.au}
\orcid{0000-0001-7243-3225}
\affiliation{%
  \institution{RMIT University}
  \city{Melbourne}
  \country{Australia} 
}
\affiliation{
  \institution{University of New South Wales}
  \city{Sydney}
  \country{Australia} 
}

\author{Hongyu Zhang}
\orcid{0000-0002-3063-9425}
\email{hongyu.zhang@newcastle.edu.au}
\affiliation{%
  \institution{The University of Newcastle}
  \city{Newcastle}
  \country{Australia} 
}

\author{Huong Ha}
\orcid{0000-0003-2463-7770}
\email{huong.ha@rmit.edu.au}
\affiliation{%
  \institution{RMIT University}
  \city{Melbourne}
  \country{Australia} 
}

\author{Flora Salim}
\orcid{0000-0002-1237-1664}
\email{flora.salim@unsw.edu.au}
\affiliation{%
  \institution{University of New South Wales}
  \city{Sydney}
  \country{Australia} 
}

\author{Xiuzhen Zhang}
\orcid{0000-0001-5558-3790}
\email{xiuzhen.zhang@rmit.edu.au}
\affiliation{%
 \institution{RMIT University}
 \city{Melbourne}
 \country{Australia} 
}

\renewcommand{\shortauthors}{Pham et al.}

\begin{abstract}
Root cause analysis (RCA) for microservice systems has gained significant attention in recent years. However, there is still no standard benchmark that includes large-scale datasets and supports comprehensive evaluation environments. In this paper, we introduce RCAEval, an open-source benchmark that provides datasets and an evaluation environment for RCA in microservice systems. First, we introduce three comprehensive datasets comprising 735 failure cases collected from three microservice systems, covering various fault types observed in real-world failures. Second, we present a comprehensive evaluation framework that includes fifteen reproducible  baselines covering a wide range of RCA approaches, with the ability to evaluate both coarse-grained and fine-grained RCA. We hope that this ready-to-use benchmark will enable researchers and practitioners to conduct extensive analysis and pave the way for robust new solutions for RCA of microservice systems.  
\end{abstract}



\maketitle

\input{1.introduction}

\input{2.datasets}

\input{3.framework}

\input{4.results}

\section{Conclusion} \label{sec:conclusion}

This paper presents RCAEval, which includes three datasets collected from three microservice systems covering 735 failure cases and 11 fault types, along with a comprehensive evaluation framework with a wide range of reproducible RCA baselines. To the best of our knowledge, this is the first comprehensive benchmark for RCA of microservices. We hope that this benchmark will be useful for industry practitioners and academic researchers in the field. 

\begin{acks}
This research was supported by the Australian Research Council Discovery Project (DP220103044). We would like to thank RMIT Race Hub for providing the computing support via the RMAS scheme. We also extend our gratitude to Dr.~Kien Do for his generous support through AWS Cloud Credit. 
\end{acks}

\bibliographystyle{ACM-Reference-Format}
\bibliography{reference}
\end{document}

%% file: 1.introduction.tex
\section{Introduction}

Root cause analysis (RCA) for microservice systems is an important problem that has been studied recently, as failures are inevitable, and ensuring the reliability of microservice systems is critical. RCA aims to analyse the available telemetry data (i.e., metrics, logs, and traces) of the system during failure periods to identify the root cause service and root cause indicators (e.g., specific metrics or logs pointing to the root cause). This field has gained significant attention recently~\cite{lee2023eadro, yu2023nezha, pham2024baro, pham2024rcaeval, pham2024root}.
However, there is still no standard benchmark that includes large-scale datasets and a comprehensive evaluation framework~\cite{cheng2023ai}. This limitation leads to inconsistent evaluations in RCA studies, hindering understanding and impeding progress in the field~\cite{cheng2023ai}. For example, existing studies~\cite{Azam2022rcd, Li2022Circa, run2024aaai} typically evaluate their methods on only 1-2 systems with 2-3 fault types. Eadro~\cite{lee2023eadro} uses a dataset with an unrealistic load (2-3 requests per second). Existing open-source RCA resources also suffer from several limitations. For example, PyRCA~\cite{liu2023pyrca} offers only a limited set of metric-based RCA methods and relies on synthetic datasets. The AIOps 2020 dataset~\cite{li2022constructing} contains failures with metrics and traces but omits valuable log information. Pham~et.~al.~\cite{pham2024rcaeval} evaluates only on metric-based RCA methods. As a result, existing resources are often inadequate for benchmarking purposes, hampering the development of new RCA approaches, see Table \ref{tab:intro}.


\begin{table}[ht]
\vspace{-0.25cm}
\caption{Comparison of studies.}
\vspace{-0.25cm}
\label{tab:intro}
\resizebox{\columnwidth}{!}{%
\begin{tabular}{l|l|c|c|c}
\hline
\textbf{Study} & \textbf{Fault Types} & \textbf{Metric} & \textbf{Log} & \textbf{Trace} \\ \hline \hline
PyRCA~\cite{liu2023pyrca} & Synthetic & \checkmark & - & - \\ \hline
AIOps 2020~\cite{li2022constructing} & Resource, Network & \checkmark & - & \checkmark \\ \hline
Pham et al.~\cite{pham2024rcaeval} & Resource, Network & \checkmark      & - & - \\ \hline 
\textbf{RCAEval (ours)} & Resource, Network, Code-level  & \checkmark & \checkmark   & \checkmark \\ \hline
\end{tabular}%
}
\vspace{-0.34cm}
\end{table}

\begin{figure*}
\vspace{-0.5cm}
\centering
\includegraphics[width=\linewidth]{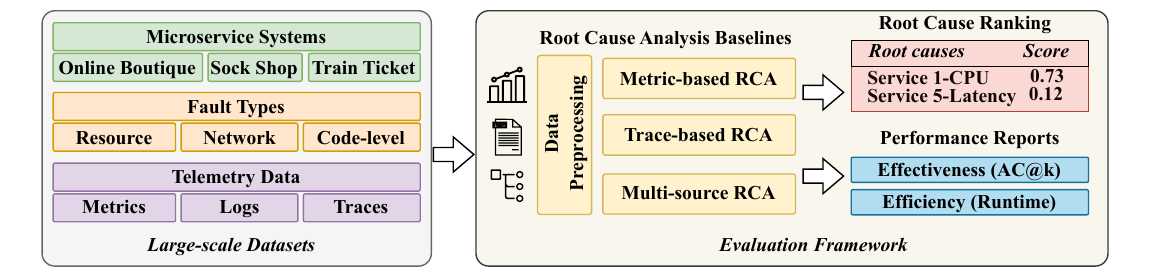}
\vspace{-0.8cm}
\caption{Overview of the RCAEval benchmark. } 
\label{fig:enter-label}
\vspace{-0.3cm}
\end{figure*}

In this work, we introduce RCAEval, a benchmark including three datasets and a comprehensive evaluation environment. First, our datasets include~735 failure cases collected from three systems, covering 11 fault types observed in real-world failures. We collected multi-source telemetry data (i.e., metrics, logs, and traces), supporting a variety of RCA approaches (e.g., metric-based, multi-source RCA). Second, we open-source our evaluation framework, which includes several reproducible baselines. The prior version of this framework was used to evaluate metric-based RCA~\cite{pham2024rcaeval}. In this work, we have upgraded it to support trace-based and multi-source RCA. We also provide preliminary experiments, highlighting the need for further investigation on a robust RCA approach. We have open-sourced our benchmark, including the datasets and the evaluation framework at~\href{https://github.com/phamquiluan/RCAEval}{\textbf{https://github.com/phamquiluan/RCAEval}}.

%% file: 2.datasets.tex
\vspace{-0.1cm}
\section{Background \& Related Work} 
\label{sec:background}

In microservice systems, \textit{failures} refer to a service's inability to perform its intended functions, while \textit{faults} represent the underlying causes of such failures (e.g., memory leaks). \textit{Root cause analysis (RCA)} seeks to pinpoint the root causes of failures by analyzing multi-source telemetry data (i.e., \textit{metrics}, \textit{logs}, and \textit{traces})~\cite{pham2024baro, pham2024rcaeval, pham2024root}. 

A major limitation in this field is the absence of a reproducible and open-source public benchmark for evaluating RCA in practical scenarios. Most RCA studies evaluate their methods using limited faults on limited systems~\cite{pham2024rcaeval, cheng2023ai}. For example, some works~\cite{lee2023eadro, yu2023nezha, Azam2022rcd} inject 2-3 faults into 1-2 systems, resulting in limited datasets. Others assess their solutions using private data, such as AWS~\cite{Azam2022rcd} or Oracle~\cite{Li2022Circa}.  This lack of transparent and reproducible resources hinders progress and prevents fair evaluation of new RCA approaches.

There have been some related works that introduce datasets or evaluation frameworks, but all of them suffer from several limitations, see Table~\ref{tab:intro}. PyRCA~\cite{liu2023pyrca} from Salesforce supports only metric-based RCA and relies on synthetic datasets. Our prior work~\cite{pham2024rcaeval} demonstrates that performance on synthetic datasets often fails to reflect RCA performance on real systems. Li et al.\cite{li2022constructing} introduced datasets with metrics and traces 
on private systems but omitting logs and did not provide a benchmarking framework. 
To address these limitations, in this study, we provide a benchmark consisting of three RCA datasets and an open-source evaluation environment.


\section{Datasets} 
\label{sec:dataset}

RCAEval benchmark includes \textbf{three datasets}: RE1, RE2, and RE3, designed to comprehensively support benchmarking RCA in microservice systems. Together, our three datasets feature 735 failure cases collected from three microservice systems (described in Section~\ref{sec:microservice-system}) and including 11 fault types (described in Section~\ref{sec:fault-type}). Each failure case also includes annotated root cause service and 
root cause indicator (e.g., specific metric or log indicating the root cause). The statistics of the datasets are presented in Table~\ref{tab:dataset}.

\noindent \textbf{RE1 Dataset.} The RE1 dataset, introduced in our prior work on metric-based RCA~\cite{pham2024rcaeval}, contains 375 failure cases collected from three microservice systems (125 cases per system). These cases combine five fault types across five services, and five repetitions per fault-service pair. The RE1 dataset exclusively contains metrics data, supporting the development of metric-based RCA methods. The fault types in RE1 include CPU, MEM, DISK, DELAY, LOSS (see Section~\ref{sec:fault-type}). The number of metrics ranges from 49 to 212, depending on the system size, with smaller systems (e.g., Online Boutique, Sock Shop) having fewer metrics compared to larger system (Train Ticket). This dataset does not include logs or traces. 

\noindent \textbf{RE2 Dataset.} The RE2 dataset, newly collected for this study, supports the development of multi-source RCA methods. It includes 270 failure cases 
(90 cases per system), combining six fault types across five services, and three repetitions per fault-service pair. RE2 provides  multi-source telemetry data, including metrics, logs, and traces. The number of metrics ranges from 77 to 327 per failure case. Each system generates a substantial volume of logs from (8.6 to 26.9 million lines), and traces (39.6 to 76.7 million traces). The fault types include those in RE1 and an additional SOCKET fault.

\noindent \textbf{RE3 Dataset.} The RE3 dataset, also newly collected, focuses on supporting multi-source RCA methods with the ability to diagnose code-level faults. It has 90 failure cases (30 per system), involving code-level faults. The fault types in RE3 are F1, F2, F3, F4, F5 (see Section~\ref{sec:fault-type}). Like RE2, RE3 includes multi-source telemetry data (metrics, logs, and traces). This dataset emphasizes diagnosing code-level faults through telemetry data, e.g., leveraging stack traces in logs or response code in traces to pinpoint root causes, making it invaluable for advancing multi-source RCA methods.

\begin{table}[ht] 
\vspace{-0.2cm}
\caption{Statistics of the RCAEval datasets.}
\vspace{-0.3cm}
\label{tab:dataset}
\setlength\tabcolsep{1pt}
\resizebox{\columnwidth}{!}{%
\begin{tabular}{l|c|c|c|c|c|c}
\hline
\textbf{Dataset} & \textbf{Systems} & \textbf{Fault Types} & \textbf{Cases} & \textbf{Metrics} & \textbf{Logs (millions)} & \textbf{Traces (millions)} \\ \hline
RE1~\cite{pham2024rcaeval} & 3 & 3 Resource, 2 Network & 375 & 49–212 & N/A & N/A \\ \hline
RE2 & 3 & 4 Resource, 2 Network & 270 & 77–376 & 8.6–26.9 & 39.6–76.7 \\ \hline
RE3 & 3 & 5 Code-level & 90 & 68–322 & 1.7–2.7 & 4.5–4.7 \\ \hline
\end{tabular}%
}
\vspace{-0.3cm}
\end{table}




\vspace{-0.2cm}
\subsection{Microservice Systems} \label{sec:microservice-system}

We collect our three datasets from three microservice systems, ranging from 12 to 64 services. These systems are used in our previous works for evaluating RCA methods~\cite{pham2024rcaeval, pham2024baro}.

\noindent \textbf{1) Online Boutique.} The Online Boutique system~\cite{ob}, developed by Google, consists of 12 services forming an e-commerce application where users can browse, add, and purchase items. The services communicate with each other using the gRPC protocol.
 
\noindent \textbf{2) Sock Shop.} The Sock Shop system~\cite{sockshop}, developed by Weaveworks, is a sock-selling e-commerce application comprising 15 services that communicate via HTTP requests.
 
\noindent \textbf{3) Train Ticket.} Train Ticket~\cite{tt} is a ticket booking system with~64 services, featuring both synchronous and asynchronous communication. Compared to Sock Shop and Online Boutique, Train Ticket has more complex call chains. To the best of our knowledge, it is the largest benchmark microservice systems.

While no single system can fully capture the diversity of real-world environments, the developers of these systems have intentionally included diverse features, such as multiple programming languages (e.g., Java, Go, Python, C\#) and communication protocols 
(e.g., HTTP, gRPC), to emulate real-world complexity.

\vspace{-0.2cm}
\subsection{Fault Types} \label{sec:fault-type}

Our three datasets consist of 11 fault types (4~resource faults, 2~network faults, and 5~code-level faults). In this section, we describe these faults and the way we introduce them into the microservice systems. The RE1 dataset, which includes 3 resource faults and 2 network faults, was used in our previous metric-based RCA works\cite{pham2024rcaeval}. In this study, we introduce two additional datasets, RE2 and RE3, which include one additional resource fault and 5 new code-level faults, covering a broader range of faults commonly found in open-source projects~\cite{cotroneo2019bad}. To the best of our knowledge, our datasets are the first to cover code-level faults for RCA in microservice systems.

\noindent \textbf{1) Resource Faults.} We introduce four resource faults into the running container (i.e., service instance) using stress-ng: CPU hog~(\textbf{CPU}), Memory leak (\textbf{MEM}), Disk stress (\textbf{DISK}), and Socket stress (\textbf{SOCK}). Symptoms of resource faults may include observable changes in resource usage of co-located containers, increased latency, and time-out requests. The system may crash when resources are severely constrained. The root cause indicator for these faults is the metric specifying resource usage (e.g., for a CPU hog, the root cause indicator is the container's CPU usage metric).

\noindent \textbf{2) Network Faults.} We use traffic control (tc) to intercept the network packets of the running container, introducing delay variations~(\textbf{DELAY}) or randomly dropping packets (\textbf{DROP}). Symptoms of network faults may include increased latency metrics and error response codes in traces/metrics of the affected service. The root cause indicator for a DELAY fault is the latency metric, while for a LOSS fault, it is the metric showing failed requests and/or error response codes in the traces of the corresponding container.

\noindent \textbf{3) Code-Level Faults.} We modify the source code of random services to introduce five bugs commonly found in open-source projects~\cite{cotroneo2019bad}: Incorrect parameter values (\textbf{F1}), Missing parameters~(\textbf{F2}), Missing Function Call (\textbf{F3}), Incorrect Return Values (\textbf{F4}), and Missing Exception Handlers (\textbf{F5}). Symptoms of code-level faults may include increased failed requests, error response codes in traces, higher latency, and stack traces emitted in logs. The root cause of code-level faults is determined using the stack traces in logs of the corresponding service, which indicate the faulty line of code. If stack traces are unavailable, the root cause indicator may be derived from error logs or response codes of the affected service.

\subsection{Telemetry Data Collection Process}

\begin{figure}
\vspace{-10pt}
\includegraphics[width=\linewidth]{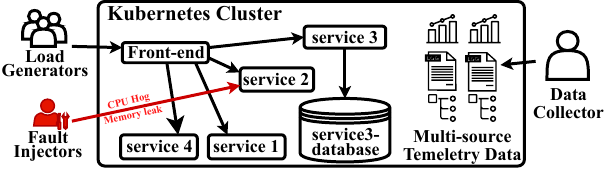}
\vspace{-26pt}
\caption{Illustration of our data collection setup.}
\label{fig:experiment-setup}
\vspace{-10pt}
\end{figure}

We deploy three microservice systems to Kubernetes clusters and generate a random load of 10–200 requests per second across all services. We use standard, well-known open-source tools to monitor and collect telemetry data. To gather metrics, we use Prometheus, cAdvisor, and Istio to monitor and collect both application-level and resource-level metrics. For logs, we use Vector and Loki to gather logs from all service instances and store them in Elasticsearch. Traces are collected using Jaeger and sent to Elasticsearch for storage, see Fig.~\ref{fig:experiment-setup}. We allow the microservice systems to run normally 
to collect normal telemetry data. Then, we inject a fault into a randomly selected running service and collect the abnormal telemetry data. 
To ensure data quality, we engaged a DevOps engineer with five years of experience in microservices to assist with system deployment, data collection, and data verification. 

\begin{table}[t]
\vspace{-13pt}
\centering
\begin{minipage}[t]{0.13\textwidth}
\centering
\caption{Metrics.} \label{tab:metrics}
\vspace{-10pt}
\resizebox{\textwidth}{!}{%
\begin{tabular}{|c|c|c|}
\hline
\textbf{time} & \textbf{cart\_cpu} & \textbf{cart\_mem} \\ \hline
17336 & 0.216 & 0.352 \\ 
17337 & 0.115 & 0.401 \\ 
17338 & 0.116 & 0.386 \\  
17339 & 0.118 & 0.398 \\  \hline
\end{tabular}%
}
\end{minipage}
\hfill
\begin{minipage}[t]{0.15\textwidth}
\centering
\caption{Logs.} \label{tab:logs}
\vspace{-10pt}
\resizebox{\textwidth}{!}{%
\begin{tabular}{|c|l|l|}
\hline
\textbf{time} & \textbf{service} & \textbf{message} \\ \hline
17336 & cart & GetCart called... \\ 
17337 & currency & Getting values...  \\ 
17338 & frontend & request complete.  \\ 
17339 & frontend & request started.  \\ \hline
\end{tabular}%
}
\end{minipage}
\hfill
\begin{minipage}[t]{0.18\textwidth}
\centering
\caption{Traces.}\label{tab:traces}
\vspace{-10pt}
\resizebox{\textwidth}{!}{%
\setlength\tabcolsep{1pt}
\begin{tabular}{|l|l|l|l|l|}
\hline
\textbf{time} & \textbf{id} & \textbf{service} & \textbf{operation} & \textbf{duration} \\ \hline
17336 & cf8b.. & frontend & GetCurrencies & 497 \\ 
17337 & 60cf.. & currency & Convert & 102 \\ 
17338 & 4a93.. & frontend & GetProduct & 1310 \\
17338 & fe23.. & product.. & ListProducts & 56 \\ \hline
\end{tabular}%
}
\end{minipage}
\vspace{-11pt}
\end{table}

\subsection{Data Format} \label{sec:data-statistic}

The raw telemetry data collected is stored as CSV files. The key structures are presented in Tables~\ref{tab:metrics}, \ref{tab:logs}, and \ref{tab:traces}. A complete dataset with full structure can be downloaded from our GitHub repository. Metrics are stored as time series data, with each row corresponding to a timestamp at which the metrics were collected. 
Logs for a failure case are stored in a single CSV file, with each row containing the timestamp, the service name, and the corresponding log message. Similarly, traces for a failure case are stored in a CSV file, where each row includes the timestamp, trace\_id, span\_id, service name, operation, duration, and response code if available.

%% file: 3.framework.tex
\begin{table*}[ht]
\vspace{-6pt}
\centering
\caption{RCA performance of eight baselines on the Train Ticket system of the RE2 dataset, across six fault types.}
\vspace{-9pt}
\label{tab:prelim-1}
\resizebox{\textwidth}{!}{%
\setlength\tabcolsep{1pt}
\begin{tabular}{c|l|rrr|rrr|rrr|rrr|rrr|rrr|rrr}
\hline
\multirow{2}{*}{\begin{tabular}[c]{@{}c@{}}Data\\ Source\end{tabular}} & \multicolumn{1}{c|}{\multirow{2}{*}{Method}} & \multicolumn{3}{c|}{\textbf{CPU}} & \multicolumn{3}{c|}{\textbf{MEM}} & \multicolumn{3}{c|}{\textbf{DISK}} & \multicolumn{3}{c|}{\textbf{SOCKET}} & \multicolumn{3}{c|}{\textbf{DELAY}} & \multicolumn{3}{c|}{\textbf{LOSS}} & \multicolumn{3}{c}{\textbf{AVERAGE}} \\ \cline{3-23} 
 & \multicolumn{1}{c|}{} & \multicolumn{1}{c}{\textit{AC@1}} & \multicolumn{1}{c}{\textit{AC@3}} & \multicolumn{1}{c|}{\textit{Avg@5}} & \multicolumn{1}{c}{\textit{AC@1}} & \multicolumn{1}{c}{\textit{AC@3}} & \multicolumn{1}{c|}{\textit{Avg@5}} & \multicolumn{1}{c}{\textit{AC@1}} & \multicolumn{1}{c}{\textit{AC@3}} & \multicolumn{1}{c|}{\textit{Avg@5}} & \multicolumn{1}{c}{\textit{AC@1}} & \multicolumn{1}{c}{\textit{AC@3}} & \multicolumn{1}{c|}{\textit{Avg@5}} & \multicolumn{1}{c}{\textit{AC@1}} & \multicolumn{1}{c}{\textit{AC@3}} & \multicolumn{1}{c|}{\textit{Avg@5}} & \multicolumn{1}{c}{\textit{AC@1}} & \multicolumn{1}{c}{\textit{AC@3}} & \multicolumn{1}{c|}{\textit{Ag@5}} & \multicolumn{1}{c}{\textit{AC@1}} & \multicolumn{1}{c}{\textit{AC@3}} & \multicolumn{1}{c}{\textit{Avg@5}} \\ \hline \hline
\multirow{5}{*}{Metric} & BARO & 0.47 & 0.8 & 0.72 & 0.93 & 1 & 0.99 & 1 & 1 & 1 & 0.6 & 0.87 & 0.83 & 0.47 & 0.67 & 0.63 & 0.53 & 0.6 & 0.64 & 0.67 & 0.82 & 0.8 \\
 & CausalRCA & 0.4 & 0.63 & 0.59 & 0.1 & 0.27 & 0.24 & 0.43 & 0.83 & 0.75 & 0.23 & 0.5 & 0.45 & 0.13 & 0.23 & 0.21 & 0.03 & 0.37 & 0.33 & 0.22 & 0.47 & 0.43 \\
 & CIRCA & 0.27 & 0.27 & 0.28 & 0.47 & 0.73 & 0.68 & 0.53 & 0.67 & 0.64 & 0.27 & 0.53 & 0.52 & 0.2 & 0.27 & 0.28 & 0.2 & 0.33 & 0.35 & 0.32 & 0.47 & 0.46 \\
 & MicroCause & 0.19 & 0.44 & 0.4 & 0 & 0.09 & 0.07 & 0.4 & 0.4 & 0.4 & 0 & 0.17 & 0.15 & 0 & 0.22 & 0.13 & 0 & 0 & 0.07 & 0.1 & 0.22 & 0.2 \\ 
 & RCD & 0.13 & 0.13 & 0.16 & 0.07 & 0.07 & 0.07 & 0 & 0.07 & 0.05 & 0.13 & 0.33 & 0.29 & 0.13 & 0.13 & 0.15 & 0.07 & 0.07 & 0.07 & 0.09 & 0.13 & 0.13 \\ \hline
\multirow{2}{*}{Trace} & MicroRank & 0.21 & 0.43 & 0.34 & 0.25 & 0.38 & 0.33 & 0 & 0.36 & 0.27 & 0.3 & 0.4 & 0.36 & 0.08 & 0.31 & 0.23 & 0.14 & 0.36 & 0.3 & 0.16 & 0.37 & 0.31 \\
 & TraceRCA & 0.64 & 0.79 & 0.74 & 0.63 & 0.88 & 0.83 & 0.64 & 0.71 & 0.74 & 0.6 & 0.8 & 0.76 & 0.85 & 0.85 & 0.88 & 0.57 & 0.71 & 0.67 & 0.66 & 0.79 & 0.77 \\ \hline
\multirow{4}{*}{\begin{tabular}[c]{@{}c@{}}Multi-\\ Source\end{tabular}} 
 & BARO & 0.47 & 0.8 & 0.75 & 0.93 & 1 & 0.99 & 1 & 1 & 1 & 0.6 & 0.8 & 0.79 & 0.47 & 0.67 & 0.61 & 0.67 & 0.67 & 0.71 & 0.69 & 0.82 & 0.81 \\ 
 & CIRCA & 0 & 0.07 & 0.09 & 0.07 & 0.13 & 0.21 & 0 & 0.07 & 0.09 & 0.07 & 0.13 & 0.16 & 0.07 & 0.07 & 0.07 & 0.13 & 0.2 & 0.17 & 0.06 & 0.11 & 0.13 \\
 & PDiagnose & 0.6 & 0.87 & 0.81 & 0.4 & 0.47 & 0.48 & 0.33 & 0.73 & 0.69 & 0.33 & 0.67 & 0.6 & 0.87 & 0.87 & 0.87 & 0.33 & 0.6 & 0.57 & 0.48 & 0.7 & 0.67 \\ 
 & RCD & 0.17 & 0.84 & 0.72 & 0 & 0.44 & 0.39 & 0.07 & 0.76 & 0.62 & 0.21 & 0.77 & 0.66 & 0.05 & 0.28 & 0.25 & 0.07 & 0.75 & 0.6 & 0.1 & 0.64 & 0.54 \\
 \hline
\end{tabular}%
}
\vspace{-10pt}
\end{table*}

\section{Evaluation Framework} \label{sec:framework}

To ensure the comprehensiveness of RCAEval, we also provide an evaluation framework as an open-source library alongside our datasets. Our evaluation framework includes fifteen baselines covering a wide range of state-of-the-art RCA approaches and offers functionalities for data processing and benchmark evaluation at both coarse-grained and fine-grained levels. The RCAEval evaluation framework is an extension of our previous work \cite{pham2024rcaeval}, which focused on metric-based RCA and coarse-grained RCA. In this work, we expand it by incorporating trace-based and multi-source RCA baselines. RCAEval is released as an open-source library and can be installed via PyPI. Comprehensive documentation on installing and using the framework with our datasets, as well as guidance on extending it with new methods and datasets, is available on our GitHub repository. The documentation also includes basic usage examples and detailed instructions for ensuring reproducibility. 

\subsection{Evaluation Baselines}


Our evaluation framework features 15 baselines covering a variety of state-of-the-art RCA methods. \textbf{Metric-based RCA baselines} include causal inference-based methods such as RUN, CausalRCA, CIRCA, RCD, MicroCause, EasyRCA, MSCRED, as well as non-causal methods such as BARO, and $\epsilon$-Diagnosis~\cite{pham2024rcaeval, liu2023pyrca, Li2022Circa, Azam2022rcd}. \textbf{Trace-based RCA baselines} include TraceRCA and MicroRank~\cite{yu2021microrank, dan2021tracerca}. \textbf{Multi-source RCA baselines} include PDiagnose, multi-source BARO, multi-source RCD, multi-source CIRCA~\cite{pham2024baro, Li2022Circa, Azam2022rcd}. For baselines like RUN, CausalRCA, CIRCA, RCD, MicroCause, EasyRCA, MSCRED, BARO, $\epsilon$-Diagnosis, MicroRank, and TraceRCA, we adapt their available implementations and use the default hyperparameter settings recommended in their respective papers. We verified their correctness by reproducing the results presented in the original and related studies. For multi-source BARO, multi-source RCD, and multi-source CIRCA, we updated their source code to handle time series data from logs and traces. For PDiagnose, we follow previous works \cite{yu2023nezha, zhang2023diagfusion, hou2021pdiagnose} to implement it since its source code is unavailable. 
Previous works such as PyRCA~\cite{liu2023pyrca} and Pham et al.~\cite{pham2024root} offer only a limited set of metric-based RCA methods, while our framework provides a more comprehensive set of baselines by also including trace-based and multi-source RCA methods.

\vspace{-0.1cm}
\subsection{Evaluation Metrics}

We support evaluation at both the coarse-grained level (i.e., root cause service) and the fine-grained level (i.e., root cause indicator). The evaluation script executes the analysis and stores the results in a report file. We currently support two standard metrics~\cite{pham2024rcaeval, pham2024baro}: $AC@k$ and $Avg@k$ to measure the RCA performance. Given a set of failure cases A, $AC@k$ is calculated as $AC@k = \frac{1}{|A|} \sum\nolimits_{a\in A}\frac{\sum_{i<k}R^a[i]\in V^a_{rc}}{min(k, |V^a_{rc}|)}$, where $R^a[i]$ is the $i$th ranking result for the failure case $a$ by an RCA method, and $V^a_{rc}$ is the true root cause set of case $a$. $AC@k$ represents the probability the top $k$ results of the given method include the true root causes. Its values range from $0$ to $1$, with higher values indicating better performance. $Avg@k$, which shows the overall RCA performance, is measured as $Avg@k = \frac{1}{k}\sum_{j=1}^k AC@j$.

%% file: 4.results.tex
\section{Preliminary Experiments} \label{sec:experiment}

We conduct preliminary experiments on our benchmark to evaluate the performance of existing baselines on the collected datasets, highlighting both the potential and challenges in the field. Due to space constraints, we select 11 baselines: 5 metric-based RCA methods (BARO, CausalRCA, CIRCA, MicroCause), 2 trace-based RCA methods (MicroRank, TraceRCA), and 4 multi-source RCA methods (PDiagnose, multi-source CIRCA, multi-source RCD, multi-source BARO). These methods are used to diagnose 4 resource faults (CPU, MEM, DISK, SOCK) and 2 network faults (DELAY, LOSS) using data collected from the Train Ticket system in the RE2 dataset. The RCA performance is evaluated using the AC@1, AC@3, and Avg@5 metrics, with coarse-grained results presented in Table~\ref{tab:prelim-1}. A demonstration of diagnosing root causes for code-level faults (e.g., F1 to F5) is available on our GitHub repository.

Our preliminary results show that there is still ample room for further improvement. Existing methods mostly obtain moderate results. For example, CIRCA and RCD obtain the best average Avg@5 score of 0.46 and 0.54, respectively. Notably, BARO shows encouraging results when obtaining high accuracy in diagnosing the resource fault (e.g. DISK), however, it shows limitations when dealing with network faults (e.g. DELAY, LOSS). Hence, we believe further research is needed to develop a holistic RCA solution.